\documentclass[runningheads]{llncs}
\usepackage[T1]{fontenc}
\usepackage{graphicx}
\usepackage{csquotes}
\begin{document}
\title{Justifications for Democratizing AI Alignment and Their Prospects}
%
%
\author{Andre Steingrüber\inst{1} \and
Kevin Baum\inst{1,2}}
\authorrunning{A. Steingrüber and K. Baum}
%
\institute{German Research Center for Artificial Intelligence (DFKI), 66123 Saarbücken, Germany \and
Center for European Research in Trusted Artificial Intelligence (CERTAIN), 66123 Saarbrücken, Germany}
\maketitle              
\begin{abstract}
The AI alignment problem comprises both technical and normative dimensions. While technical solutions focus on implementing normative constraints in AI systems, the normative problem concerns determining what these constraints should be. This paper examines justifications for democratic approaches to the normative problem---where affected stakeholders determine AI alignment---as opposed to epistocratic approaches that defer to normative experts. We analyze both instrumental justifications (democratic approaches produce better outcomes) and non-instrumental justifications (democratic approaches prevent illegitimate authority or coercion). We argue that normative and metanormative uncertainty create a justificatory gap that democratic approaches aim to fill through political rather than theoretical justification. However, we identify significant challenges for democratic approaches, particularly regarding the prevention of illegitimate coercion through AI alignment. Our analysis suggests that neither purely epistocratic nor purely democratic approaches may be sufficient on their own, pointing toward hybrid frameworks that combine expert judgment with participatory input alongside institutional safeguards against AI monopolization.
\end{abstract}
\section{Democratic approaches to the normative problem of AI alignment}
The AI alignment problem consists of two sub-problems: a technical problem and a normative problem \cite[p. 412-13]{Gabriel2020}. The technical problem is a question of machine ethics and requires us to find algorithmic implementations of normative constraints that effectively regulate the behaviour of AI systems. The normative, primarily philosophical problem, on the other hand, requires us to determine what these constraints should be. In this paper we will be concerned with the normative problem and potential justifications for solving it democratically.

There are two main ways to determine the content of an AI's normative constraints. The first is to take a top-down approach and determine the content of normative constraints by consulting normative philosophical theories or by deferring to people identified as normative experts, see \cite{Askell2021,Bai2022a,Bai2022b,Kim2021,Riesen2025}. Secondly, one can take a bottom-up approach, letting relevant stakeholders---for instance, all those affected by an AI's alignment (directly or indirectly)---determine the content of its normative constraints, see \cite{Awad2018,Christiano2017,Gabriel2025,Jiang2022,Ziegler2019}. Call the former \enquote{epistocratic approaches} and the latter \enquote{democratic approaches}.\footnote{Pluralistic approaches would combine epistocratic and democratic elements to determine an AI's normative constraints. As we will explain in the next section, the task of producing normative constraints for an AI can be broken up into three steps: For every scenario, we need to (i) identify the relevant reasons, (ii) measure the relative strength of these reasons, and (iii) aggregate the relevant reasons to form overall deontic verdicts that formulate the AI's normative constraints. Approaching these three steps in a pluralistic fashion, one can partition the set of scenarios and handle one subset epistocratically and the other subset democratically, or one can let epistocratic approaches take care of certain steps of the procedure and let democratic approaches do the remaining steps; or, alternatively, one can involve both epistocratic and democratic approaches in a single step, e.g., identifying the relevant reasons by eliciting people's judgements on the matter and then letting experts add missing reasons they consider important, or letting them veto against particular reasons.}

It is important to note, however, that democratic approaches may be democratic in name only, as they may fail to be genuinely democratic depending on the procedure being used to determine the normative constraints from the input of the people. Some procedures, if done correctly, are apt to be democratic---e.g., voting, sortition, or deliberation---others, like a knockout tournament in bowling or a debating contest, much less so.

Schuster and Kilov \cite{Schuster2025} argue that current proposals for democratic approaches all invoke procedures that fail to be democratic. However, we believe this is based on a misunderstanding. Let us clarify, because this misunderstanding comes up often in the alignment literature: The approaches that Schuster and Kilov discuss are crowdsourcing normative judgement, reinforcement learning from human feedback (RLHF), and constitutional AI. Yet, these three techniques should \textit{not} be understood as solutions to the normative problem, but rather as solutions to the \textit{technical} problem.\footnote{See, e.g., also \cite[p. 414]{Gabriel2020}, \cite[pp. 12-13]{Huang2025}, and \cite[p. 2672]{Kneer2025}, for instances where RLHF and constitutional AI are treated as solutions to the normative problem.} In the case of crowdsourcing and RLHF, normative constraints are given implicitly in the form of a large number of individual human normative judgements \cite{Christiano2017,Jiang2022}, and with constitutional AI, normative constraints are given explicitly in the form of a list of principles formulated in natural language \cite{Bai2022b}. The primary aim of these techniques is to implement normative constraints in AI systems and not to determine what the normative constraints should be.

Crowdsourcing, RLHF and constitutional AI are all compatible both with epistocratic and democratic approaches to the normative problem. Although constitutional AI may sound like it particularly lends itself to epistocratic approaches, and crowdsourcing and RLHF like they are especially suited for democratic approaches, there is no such association for any of these techniques. In general, it is possible to determine the list of principles necessary for constitutional AI both via an epistocratic or a democratic approach; either we consult normative theories to derive a constitution, or we ask all affected people to determine one. Likewise for RLHF and crowdsourcing, what the input data should be can either be decided by normative experts, or by the affected public. We must be careful to distinguish between the procedures invoked to solve the normative problem and techniques used to tackle the technical problem.

Having made this clarification, let's return to the two approaches to the normative problem. Some authors argue that democratic approaches, if they are actually democratic, should be favoured over epistocratic ones because they allow us to avoid putatively morally undesirable aspects of epistocratic approaches \cite{Gabriel2025,Huang2025,Jiang2022}. Here are some moral reasons that are claimed to disfavour epistocratic approaches:

\begin{quote}
    \enquote*{The lack of a broad, inclusive, and democratic process for determining these values can lead to AI systems that disproportionately reflect the interests of specific groups, exacerbating existing inequalities and failing to serve the broader public good.} \cite[p. 11]{Huang2025}
\end{quote}
\begin{quote}
    \enquote*{[W]e follow a bottom-up approach to Delphi for an important ethical concern: [\ldots] implementing the top-down approach would force scientists to impose their own value choices and principles in the system they build, which is not an appropriate social role for scientists alone.} \cite[p. 7]{Jiang2022}
\end{quote}
\begin{quote}
    \enquote*{[E]fforts to align AI systems with a given moral schema may lead to unjust value imposition or even domination.} \cite[p. 3]{Gabriel2025}
\end{quote}

\noindent We can group the putative reasons speaking against epistocratic approaches into two categories: Instrumental reasons against epistocratic approaches (and for democratic approaches), and non-instrumental reasons against epistocratic approaches (and for democratic approaches).\footnote{Compare \cite{Christiano2024} for this taxonomy of justifications for democratic practices.} If epistocratic approaches were to \enquote{exacerbate inequalities} or \enquote{fail to serve the public good}, they would be instrumentally worse than democratic approaches, because adopting them would have morally worse consequences. If, on the other hand, pursuing epistocratic approaches were to constitute \enquote{value imposition} or \enquote{domination}, they would be non-instrumentally worse, because they are inherently, i.e., independently of their consequences, morally objectionable.


Proponents of democratic approaches are not always explicit about what kind of justification it is that speaks in favour of their theory, and they don't always consider what theoretical resources epistocratic approaches can draw on that may undercut the justifications that are supposed to support democratic approaches. Therefore, in this paper, we want to unpack the reasons that may be used to justify democratic approaches and estimate how promising they are. We want to suggest to the proponents of democratic approaches the most promising justificatory avenues, but also point out what questions they have to answer to pave those paths. Mainly, we will focus on non-instrumental reasons, but we will also briefly touch upon the instrumental ones. Two possibilities for what might be inherently bad about epistocratic approaches will be discussed: (i) They give some people illegitimate authority over other people. (ii) Through them some people will be illegitimately coerced by others. We will argue that the latter is the more promising argumentative route for proponents of democratic approaches. However, whether it succeeds in justifying democratic approaches over epistocratic approaches depends on at least four things: that users of an AI can really be coerced through the AI's alignment; that, if users of an AI can really be coerced through the AI's alignment, this would be illegitimate; that democratic approaches can produce a democratic justification that would justify the coercion and thereby prevent illegitimate coercion; and that epistocratic approaches cannot prevent the illegitimate coercion.

However, before we turn to discuss instrumental justifications and subsequently non-instrumental justifications, we first want to consider a crucial motivation and enabling condition for democratic approaches: reasonable normative disagreement.
\section{Normative disagreement leaves a justificatory gap}
The observation that reasonable people can deeply disagree when it comes to normative matters is one motivation for proponents of the democratic approach to pursue their project \cite{Awad2018,Gabriel2025,Huang2025,Jiang2022}. It is worthwhile to consider how exactly that is so, to get clearer on what the aims and obstacles of democratic approaches are. The short version is this: The empirical fact of normative (and metanormative) disagreement makes us normatively (and metanormatively) uncertain, i.e., we are unsure what the right thing to do is (and whether there even is a uniquely right thing to do). This uncertainty eliminates what would be a straightforward justification for any potentially illegitimate state of affairs. If we were normatively (and metanormatively) certain, we could simply show that the normative constraints we are implementing are (objectively) correct. Since such a theoretical justification is unavailable, this makes it possible that, by means of an AI's normative constraints, people are given \textit{illegitimate} authority over other people, or that some people are being \textit{illegitimately} coerced by others. Let us consider this in more detail.

We are facing the normative problem under both normative and metanormative uncertainty. That is, neither are we certain what the normative ground truth is, nor are we certain whether there even is a normative ground truth and how to find out about it.

We are normatively uncertain because we can observe that reasonable people can widely diverge in their judgements about what reasons are relevant for a decision and how their strength compares to each other, and because different normative theories, like theories of normative ethics, can have very different answers to these questions \cite{MacAskill2020}. That is, our normative uncertainty is the rational response to observed intersubjective and intertheoretical normative disagreement.

Likewise for metanormative disagreement. It is the rational response to observed intersubjective and intertheoretic metanormative disagreement. We are metanormatively uncertain in at least three respects: We are uncertain whether there is a normative ground truth, i.e., whether there are robustly mind-indepen\-dent normative reasons. We are uncertain whether this normative ground truth is unique, i.e., whether normative reasons hold absolutely or only relative to some frame of reference. And we are uncertain whether and how we can have knowledge about this normative ground truth, i.e., whether there is a reliable method to identify, measure and aggregate normative reasons.

A short digression: We are deliberately speaking about normative and metanormative uncertainty in general and not just about moral and metaethical uncertainty in particular, because an AI’s normative constraints are not exhausted by moral constraints. We don’t just want to know what is morally permissible, impermissible or obligatory to do for an AI system, we want to know what is overall permissible, impermissible or obligatory \cite{Baum2025}. To properly align AI systems they have to be sensitive to normative domains other than the moral domain. Consider, e.g., that some things that are morally permissible are not legally permissible, like taking food from the supermarket’s bin, or they are not socially permissible, like talking much too loud in public spaces. To know what we and what an AI should do---Are we allowed to stand in the middle of the escalator blocking other people from walking past us?---we have to consider all relevant reasons from different relevant normative domains and weigh them against each other in order to arrive at an all-things-considered \textit{overall} deontic verdict and not just an all-things-considered \textit{moral} deontic verdict. To solve the normative problem, we thus have to: (i) \textit{identify} which practical reasons from which normative domain are relevant for a decision, (ii) \textit{measure} the strength of the relevant reasons, and (iii) \textit{aggregate} the relevant reasons according to their strength to form an all-things-considered overall reason that grounds an overall deontic verdict.

The fact that we are seeking overall reasons that play the role of overall normative constraints exacerbates our normative and metanormative uncertainty. For one, if we consider non-moral normative domains in isolation there may be even less common ground in people's judgements or conversely even more normative disagreement. Just consider the diverse social norms or legal norms that people take to hold. Which of them should we choose to align AI with? But what's more, since the reasons from different normative domains interact, this introduces an entirely new dimension of normative/metanormative uncertainty. How exactly do moral reasons, reasons of politeness, and legal reasons interact, for example? Some may tend to let legal and politeness reasons be able to take precedence over moral reasons, others will think that moral reasons always override reasons from other domains. All these uncertainties can accrue and reflect in our uncertainty about the all-things-considered overall normative reasons that, in the end, are supposed to figure as normative constraints for an AI system.

Now, how exactly do normative and metanormative uncertainty motivate democratic approaches? They do so insofar as they are necessary conditions for the possibility of the illegitimacy of authority, coercion, or relational inequality. If we were certain that some objective all-things-considered overall reason holds, then this would give us a justification to do as the reason demands. If we had decisive evidence (whatever that would look like) for the truth of a certain practical normative judgement---One ought to $\varphi$---then we would have all-things-considered theoretical reason to believe that one ought to $\varphi$ which in turn would constitute a contributory practical reason to $\varphi$. Normative certainty would therefore put us in a position to justify and thereby legitimise authority or coercion; we would be able to show that some demands are not discretionary but well founded. Conversely, this is how normative and metanormative uncertainty is a necessary condition for unjustified authority or coercion: it eliminates a sure theoretical justification that could always legitimise potentially illegitimate authority or coercion; when we don't know what the normative ground truth is, or we don't know how to find out what it is, or are not even sure that there is one, then we can't appeal to it to safely justify a potentially illegitimate state of affairs.

Normative and metanormative uncertainty thus leave us with a justificatory gap, one that democratic approaches are motivated to fill. The democratic aim is to compensate for the \textit{missing theoretical} justification of an AI's normative constraints with a \textit{political} justification. The idea being, if all people affected by an AI's normative constraints get to have a say in what these constraints are, this legitimises any potentially illegitimate authority or coercion by means of an AI's normative constraints. Preventing illegitimate authority or coercion is supposed to non-instrumentally justify democratic solutions to the normative problem. Before we consider non-instrumental justifications, however, let us briefly say a few words about instrumental justifications for democratic approaches.

\section{Instrumental justifications for democratic alignment}
Proponents of democratic approaches may justify their preferred solution to the normative problem by arguing that it, in some sense, works better than epistocratic approaches; employing democratic approaches has better consequences than not doing so. We want to mention two ways in which this might be the case, and on which defenders of the democratic approach could focus.\footnote{For a general description of both of them, see \cite{Christiano2024}.}

First, one may try to argue that if we let the people that are going to be affected by the behaviour of an aligned AI decide how it ought to be aligned, then the aligned behaviour of the AI will be better \textit{for} the people. The idea is that people know best what is good for them, or at least better than normative experts and their theories. Thus, if we let them decide, instead of only the experts, they will be better off than they otherwise would have been.

But there is still quite some argumentative work left to be done for this justification to really get off the ground. First of all, proponents of the democratic approach need to decide whether they want to read the counterfactual \enquote{If people have a say in what the normative constraints of an AI are, they would be better off (with respect to the AI's behaviour towards them) than they otherwise would have been} generically or specifically. Do they aim for a general justification of democratic approaches and want to roughly say \enquote{Typically, if people have a say in what the normative constraints of an AI are, they would be better off (with respect to the AI's behaviour towards them) than they otherwise would have been}? Or do they aim for a case-by-case justification and want to say \enquote{In this case, if people have a say in what the normative constraints of an AI are, they would be better off (with respect to the AI's behaviour towards them) than they otherwise would have been}? The latter justification is weaker but also comparatively easier to come by.

Under both readings, democratic approaches still have to argue that people would \textit{actually} be better off than they otherwise would have been. It does not seem implausible to suppose, e.g., that normative experts are subject to biases that reflect in their normative verdicts and that consequently would disadvantage certain groups of persons. If these people get to have a say, then, most likely, they will not disadvantage themselves, i.e., plausibly they would be better off. However, proponents of democratic approaches should be careful to take epistocratic approaches seriously and not to argue against straw men of them. It might be a real risk that epistocratic approaches arrive at normative constraints that are biased, but to criticise epistocratic approaches this risk has to be estimated, and additionally, democratic approaches need to show that they do not run this risk. Further, to argue successfully that it is better for people if they can democratically participate, proponents of democratic approaches have to react to objections that invoke cases where people seem to vote against their best interest; think Brexit, Trump, the climate crisis, etc.

Another possible instrumental justification we want to mention relates to the idea of the wisdom of crowds. The claim would be that, although epistocratic approaches consult the judgement of normative experts, democratic approaches are better at producing more correct results. This is an epistemic justification because the point is supposed to be that (under certain assumptions) democratic processes are better at tracking the normative facts. Typically, Condorcet's Jury Theorem is being used to argue for this point. Roughly, it states that the probability that a majority of voters choose the correct option approaches 1 as the number of voters increases. That is, the bigger the electorate, the more reliable the result of their vote \cite{Goodin2018}.

However, Condorcet's Jury Theorem relies on unrealistic assumptions. For it to hold, one needs to assume that voters' judgements are probabilistically independent of each other, and that voters are generally competent, evidenced by the fact that they are more likely to vote for the correct option than for the incorrect option. In real-life cases these assumptions are almost never satisfied. The assumptions can, however, be weakened to make the jury theorem applicable for real cases \cite{Goodin2018}. Even then though proponents of the democratic approach have to show that the weakened assumptions hold in the case of AI alignment they are considering. And they need to respond to objections, two of which we want to allude to. First, democrats have to make sure that epistocrats cannot also make use of the jury theorem, with the difference being that only normative experts comprise the electorate. And second, to employ the jury theorem one has to assume that there is an objective fact about the matter that is being voted on. In the present context of the normative problem the matter would be normative, and to assume that there is an objective fact about these matters would be a metanormative assumption. Such an assumption might be in tension with the assumption of metanormative uncertainty democratic approaches are motivated by.

Proponents of democratic approaches can argue for their proposed solution to the normative problem by resorting to these and other instrumental justifications. To reap the justificatory fruits they have to show that the advertised consequences---prudentially or epistemically better decisions---are actually achievable in the case of AI alignment, and they have to show that epistocratic approaches do not have access to the same benefits in different ways. Another way to justify democratic approaches is through non-instrumental justifications. We will turn to them now.
\section{Non-instrumental justifications for democratic alignment}
Above, we have quoted Gabriel and Keeling who worry that epistocratic approaches may lead to illegitimate \enquote{value imposition or even domination} \cite[p. 3]{Gabriel2025}. This exemplifies a non-instrumental objection to epistocratic approaches. More detailed, the worry is that, through an AI's alignment, people can be indirectly subjected to normative standards they do not subscribe to themselves. Since epistocratic approaches cannot close the justificatory gap left by normative and metanormative uncertainty, they cannot justify such subjection which makes it illegitimate. For example, if your personal AI assistant does not let you buy meat because that would be against its normative constraints, you are being subjected to normative standards to which you do not subscribe. Or, if a generative AI is uncompliant with your request to gender an email draft because its alignment forbids it to do so, other people's values are being imposed on you.

But what exactly do we mean by \enquote{value imposition}, \enquote{domination}, and \enquote{subjection}? Two possible interpretations are that they either refer to authority over the users of AI, or to the coercion of users of AI. A person with (justified) authority can issue commands or make claims that generate real reasons for action for other people \cite{Raz1986}. For example, within certain confines, teachers are typically taken to have (justified) authority over their pupils. And a person equipped with coercive power can restrict other people's freedom to act as they desire. For example, within certain confines, policemen are typically taken to wield coercive power.

What is at issue in the case of contentious AI alignment? Arguably, it is coercive power rather than authority. The question of authority would only arise if an AI system were to be deployed in a way where it issues commands or makes claims on people. The question of coercive power, on the other hand, arises as soon as an AI system restricts people's freedom to act as they desire. This can happen rather quickly. If a self-driving vehicle does not let you drive above a certain speed limit because of its normative constraints, then it doesn't command you to drive slower, it simply makes you so. And if a large language model does not let you write your text in gender-sensitive language, it makes no claim on you to not do so, it just doesn't use gender-sensitive language. We could multiply examples but the point is: Both the question of an AI's authority and its coercive power can be pertinent but we take the threat of coercion to be the more pressing and focus on it in the following.\footnote{See also Ripstein \cite{Ripstein2004}, who argues that, in general, the function of democratic justification is to legitimise coercion rather than authority.}

Proponents of democratic approaches have to argue for four things in order to be able to claim that preventing illegitimate coercion non-instrumentally justifies democratic approaches as compared to epistocratic approaches: They have to argue (i) that it is indeed possible for people to be coerced through an AI's normative constraints, (ii) that such coercion, if it is possible, would be unjustified, (iii) that democratic approaches can produce a democratic justification that would justify the coercion and thereby prevent illegitimate coercion, and (iv) that epistocratic approaches cannot prevent the illegitimate coercion. Let us make a few remarks concerning each proposition.

Is it possible for the users of an AI to be coerced through the AI's normative constraints? Two quick clarifications to begin with: First, if an AI's alignment is coercive, the primary coercer is not the AI itself but the person or organisation that defines the normative constraints. The AI is only the means of (potential) coercion. A bit more verbosely we are asking: Do the people who define the normative constraints of an AI coerce the users of the AI if (some of) the chosen normative constraints are not endorsed by the users? Second, and following from this, if users are being coerced, then only indirectly so. The primary (potential) coercers are the people who decide how the AI is aligned, but they don't compel the users directly; rather, their coercion is mediated by the AI. That an AI's alignment can only be indirectly coercive does not speak against it really being coercive. If you can be coerced by having someone limit what you can do with your bank account---think, abusive relationship---you can also be coerced by having someone limit what you can do with an AI.

More needs to be said to convincingly argue that users of an AI can be indirectly coerced by its normative constraints. Let's assume that an argument can be given for that. What would have thereby been shown is that it is \textit{in principle possible} for an AI's alignment to be coercive. What has not been shown is that there is an \textit{actual} case where it \textit{actually} is coercive. For an AI's alignment to be \textit{actually} coercive would not just depend on the relationship between the users and the aligned AI but crucially also on certain background conditions.\footnote{In a similar context Kolodny \cite[pp. 97-101]{Kolodny2023} calls these background conditions \enquote{tempering factors}.} We said that a person equipped with coercive power can restrict other people's freedom to act as they desire. Conversely this means if a person is free to act as she pleases then she is not being coerced. This matters for the case of normatively coercive AI in the following way: Say, for normative reasons, your AI assistant is noncompliant with your request to buy meat. Are you being coerced by this? Well, it depends on whether you are still free to buy meat, maybe with the help of another, differently aligned AI, or simply on your own. You wouldn't be free to do so, for example, if using the specific AI assistant were the \textit{de facto} or \textit{de jure} standard for going shopping. You would then not be able to buy meat, at least not without great opportunity cost. But if that is not the case, and without much ado you can just go and use another AI or buy the meat yourself, then you are not being coerced by the normatively noncompliant AI. You are no more being coerced than you would be when you can only buy vegan products at your local supermarket, or when you have to wear \enquote{gender appropriate clothes}---no skirts for men, no trousers for women, etc.---at your bowling club. The users of the AI, the supermarket, or the bowling club may be compelled to use them in a certain way, but this is not coercion as long as they can freely go elsewhere to use another AI, another supermarket, another bowling club.

The next point proponents of democratic approaches would have to argue for, in order to strengthen the non-instrumental justification in favour of their solution, is that coercion by means of an AI's alignment, if it is possible, would be unjustified. Two important ingredients for such an argument would presumably be the prima facie wrongness of coercion, and the fact that we are normatively and metanormatively uncertain. The first ingredient, the prima facie wrongness of coercion, may be used to establish that prima facie coercion stands in need of justification, much like murder or marital infidelity would. And the second ingredient, normative and metanormative uncertainty, may be used to establish that no straightforward theoretical justification for the AI's normative constraints is available, and consequently also not for the coercion by means of them. Thus, other things being equal, coercion by means of an AI's alignment would be unjustified.

Observe however, that there is a tension between the two ingredients. If we are normatively uncertain, how can we purchase the assumption that coercion is prima facie wrong?\footnote{We thank two reviewers for raising that point.} We are deliberately only talking about a tension and not a contradiction because proponents of democratic approaches may be able to argue that the two ingredients are consistent with each other. For example, because our normative uncertainty is not evenly distributed over all normative propositions; about some we are more certain, about some we are less certain. Coercion being prima facie wrong perhaps is of the first kind, while overall we are still normatively uncertain. Whether in this way or differently, if proponents of the democratic approach buy into both ingredients then they have to address the apparent tension between them.

Another point that needs to be addressed is if there are other justifications for (the choice of) certain normative constraints, apart from a sure theoretical justification, or the democratic justification that democratic approaches aim for. If there is another possible justification, then coercion through AI alignment would be justifiable and proponents of democratic approaches would lose the non-instrumental reason in favour of their solution. Candidates for such a justification are decision rules that are explicitly designed to deal with normative uncertainty. For example, MacAskill, Bykvist and Ord \cite{MacAskill2020} defend a rule called \enquote{maximise expected choiceworthiness}. Analogous to descriptive uncertainty they treat normative disagreement as data to approximate the correct normative constraints by assigning weights to different normative hypotheses, where different normative hypotheses determine the choiceworthiness of an outcome. Maximise expected choiceworthiness then demands to choose the action that leads to the outcome with the highest sum of weighted choiceworthiness. Although we cannot have a sure theoretical justification for any particular normative constraint, we may be able to have a practical justification to choose certain constraints over others by means of decision rules like this that deal with normative uncertainty. The practical justification we get from any such rules inherits its strength from the strength of the theoretical justification for the particular decision rule, meaning that proponents of democratic approaches either have to critique the theoretical justification for the rule, or they have to argue that the practical justification for choosing certain normative constraints we get from rules like maximise expected choiceworthiness is in general of the wrong kind.

The next point proponents of democratic approaches would have to argue for, in order to strengthen the coercion-based non-instrumental justification in favour of their solution, is that they can actually produce a democratic justification that is suited to justify potential coercion. A number of objections can be levelled against this, and would therefore have to be addressed by proponents of the democratic approach. Let us mention just two.

Any democratic approach will have to stipulate what the rules of their democratic game should be. Do people vote, if so, what's the voting procedure? Do people deliberate, if so, what are the rules of discourse? And so on. No matter what the rules end up being, for them to be recognisably democratic, they have to make normative assumptions.\footnote{For a detailed discussion of this point in context of AI ethics in general, see \cite{Baum2020}.} For example, a vote is free and equal, or a deliberation is inclusive and non-coercive. The question however is, what justifies these normative assumptions? Some of the people affected by an aligned AI may be able to reasonably reject them. But if the democratic process is not properly justified, its output will also not be. 
This, like all the other points, is not a knockdown argument, rather, it is intended to raise an issue that proponents of democratic approaches have to somehow address---namely, the bootstrapping problem of how to justify the very democratic procedures that are supposed to provide justification.

The same applies for this second objection. Given the deep normative disagreement between people, one may worry that whatever all people affected by an AI's alignment can agree on will only be the \enquote*{lowest common normative denominator}, and much too little to really prevent the threat of illegitimate coercion. The objection here is not that no democratic justification is achieved through the democratic approach, it is that the justification is too minimal to do the job it is supposed to do. The minimal output from democratic approaches has to be beefed up, in order for AI systems to really be effectively regulated, but then the threat of illegitimate coercion re-enters again.

The last point proponents of democratic approaches would have to argue for, in order to strengthen the coercion-based non-instrumental justification in favour of their solution, as opposed to the epistocratic solution, is that epistocratic approaches do not have the resources to confront the threat of illegitimate coercion through an AI's alignment. By now we have seen that it is not so clear that this is the case. Let us mention again just two reasons for thinking that epistocratic approaches can get a handle on the problem of illegitimate coercion.

We have said that the possibility of coercion depends on certain background conditions. For example, if the use of some AI system is the \textit{de facto} or \textit{de jure} standard, then people may not be free to do what they desire to do without using the AI, and then they are potentially being coerced by the AI's alignment. Conversely this means, if we make sure that for all purposes there are always multiple AIs with different alignments available, then people are free to choose the AI that does not coerce them. And in general, if we control the background conditions that are necessary for coercion, we, and proponents of epistocratic approaches in particular, can prevent the threat of illegitimate coercion by preventing the threat of coercion.

Additionally, epistocratic approaches may be able to justify their choice of normative constraints and thereby justify potential coercion by means of them. To this end, they can invoke decision rules like maximise expected choiceworthiness, that explicitly take normative uncertainty into account. If an epistocratic approach is able to appropriately justify its solution of the normative problem, and is able to transfer this justification to the coercion through an AI's alignment, then this would undercut the non-instrumental justification from coercion for democratic approaches.

%
%
\section{Conclusion}
In this paper, we have motivated democratic approaches to the normative question of AI alignment. We have discussed instrumental and non-instrumental ways of justifying them, focussing in particular on the non-instrumental justification from coercion. We have argued that proponents of democratic approaches have to argue for four propositions in order for this justification to be successful: (i) It is possible for people to be coerced through an AI's normative constraints. (ii) Such coercion, if it is possible, would be unjustified. (iii) Democratic approaches can produce a democratic justification that would justify the coercion and thereby prevent illegitimate coercion. (iv) Epistocratic approaches cannot prevent the illegitimate coercion.

We have argued that none of the four propositions is without problems and comes for free. In particular, we have argued that there are independent ways to prevent people from being coerced by means of an AI's normative constraints. Namely, we can control the background necessary conditions for coercion, e.g., we can prevent any one AI becoming the \textit{de facto} or \textit{de jure} standard for certain purposes such that people are dependent on it. If epistocratic approaches can draw on this possibility, as democratic approaches can as well, then they might be able to take the sting out of the non-instrumental justification from coercion in favour of democratic approaches.

Our analysis suggests that neither purely epistocratic nor purely democratic approaches to the normative problem may be sufficient on their own. The challenges we have identified do not eliminate the potential value of democratic participation, but rather point towards more nuanced, context-sensitive solutions. At least for some application contexts, hybrid frameworks that combine expert judgement with targeted participatory input, alongside appropriate institutional safeguards that mitigate AI monopolies and in particular AI systems that are too uniformly aligned, may therefore be the most suitable paths for addressing the normative dimensions of AI alignment. The critical question for future research is determining when and how to optimally combine epistocratic and democratic elements---specifying which aspects of the normative problem benefit from expert knowledge versus democratic input, and under what institutional conditions such hybrid approaches can succeed.

\begin{credits}
\subsubsection{\ackname} We thank two anonymous reviewers and Elijah Millgram for their feedback on versions of this paper. This work is partially funded by DFG grant 389792660 as part of TRR~248 -- CPEC, see \url{https://perspicuous-computing.science}, by the German Federal Ministry of Education and Research (BMBF) as part of the project ``MAC-MERLin'' (Grant Agreement No. 01IW24007), and from the European Regional Development Fund (ERDF) as well as the German Federal State of Saarland within the scope of the project (To)CERTAIN. 

\subsubsection{\discintname}
The authors have no competing interests to declare.
\end{credits}
%
%
%
%

\end{document}